# Holographic study of the jet quenching parameter in anisotropic systems

Luying Wang[1,a], Shang-Yu Wu[2,b]

[1] Department of Physics, Shanghai University, Shanghai 200444, People's Republic of China
[2] Department of Electrophysics, National Chiao Tung University, Hsinchu 300, Taiwan, ROC

**Abstract** We first calculate the jet quenching parameter of an anisotropic plasma with a $U(1)$ chemical potential via AdS/CFT duality. The effects of charge, anisotropy parameter, and quark motion direction on the jet quenching parameter are investigated. We then discuss the situation of an anisotropic black brane in the IR region. We study both the jet quenching parameters along the longitudinal direction and the transverse plane.

## 1 Introduction

The gauge/string duality [1–3] provides a powerful tool to analyze the dynamics of the strongly coupled quark–gluon plasma (QGP). For example, the gauge/string duality has been applied to various topics in heavy ion collisions, e.g., drag force [4–8], photon production [9–11], elliptic flow [12], and jet quenching [13–18]. Among these we mainly study anisotropy jet quenching [19–34]. The experiments conducted at the Relativistic Heavy Ion Collider (RHIC) [35,36] and at the Large Hadron Collider (LHC) [37,38] demonstrate that the QGP is anisotropic and far from equilibrium during a short period of time after the collision. After that, the system becomes locally isotropic. In Refs. [24–34], the authors constructed a new black brane solution of IIB supergravity, aiming to describe the QGP at intermediate times $\tau_{out} < \tau < \tau_{iso}$ with an intrinsically anisotropic hydrodynamical picture. This gravity solution is dual to a spatially anisotropic $\mathcal{N} = 4$ super Yang–Mills plasma at finite temperature [39,40]. A R-charged version of the anisotropic black brane was also obtained by one of us [19,20]. Those static solutions, possessing a regular, anisotropic horizon, can be viewed as a renormalization group flow from an AdS geometry in the ultraviolet to a Lifshitz-like geometry in the infrared.

In this paper, we wish to extend previous calculations of the jet quenching parameter to a more general Lifshitz-like geometry [13]. A natural realization of the AdS/CFT correspondence with Lifshitz-fixed points is described by the following anisotropic spacetime [41–48]:

$$ds^2 = r^2 \left( -dt^2 + \sum_{i=1}^{p} dx_i^2 \right) + r^{\frac{2}{z}} \sum_{i=p+1}^{d} dy_j^2 + \frac{dr^2}{r^2}, \quad (1)$$

which is invariant under the scaling transformation $(t, x_i, y_j, r) \rightarrow (\lambda t, \lambda x_i, \lambda^{\frac{1}{z}} y_j, r/\lambda)$. Note that the $y_j$-direction corresponds to the Lorentz symmetry violation and anisotropy. The simplest case with $p = 0$ represents non-relativistic fixed points with dynamical critical exponent $z$, which is very well known in quantum critical systems in condensed matter physics. Other cases with $1 \le p \le d - 1$ are interpreted as space-like Lifshitz fixed point and cannot be simply regarded as a generalization of $p = 0$. The black brane solution is given in Refs. [19,20,39], corresponding to the case $z = 3/2$. One of the main purposes of this paper is to calculate jet quenching parameters along the longitudinal direction and transverse plane.

Another purpose of this paper is to take the effect of a non-zero chemical potential into account on calculating the jet quenching parameter. In general, in the quark–gluon plasma produced in RHIC, the escaped quark is surrounded by high density quarks fluids liberated from the nucleons of the heavy ions and then some of the jets are quenched by the surrounding medium. In such a setting, the baryon density of the quark–gluon plasma is relevant and the chemical potential must be taken into account.

The structure of this paper is as follows. In Sect. 2 we build up a general setting. In Sect. 3, we consider the jet quenching in spatially anisotropic black brane background with a chemical potential. The jet quenching parameter in a spatially anisotropic black brane with a chemical potential at the quark moving along the longitudinal direction is discussed in Sect. 3.1. The cases of the transverse plane are then discussed in Sect. 3.2. The jet quenching parameter in an anisotropic black blane background as the quark is moving along the longitudinal direction is given in Sect. 4.1. The situation for the quark moving in the transverse plane is discussed in Sect. 4.2. We end with conclusions in Sect. 5.

[a] e-mail: wangluying1129@gmail.com
[b] e-mail: loganwu@gmail.com

## 2 General setup

In order to compute the jet parameter $\hat{q}$ for an ultra-relativistic quark, we need to consider the worldsheet of a string whose endpoints move at the speed of light along a given boundary direction and are separated a small distance $l$ along an orthogonal direction. As noted in [15], the jet quenching parameter depends on how these directions are oriented with respect to the longitudinal and transverse directions in the plasma. For simplicity, we refer to $z$-axis as the longitudinal direction and to the $xy$-plane as the transverse directions.

As illustrated in Fig. 1, we assume that the direction of motion is contained in the $xz$-plane, which will be denoted $Z$. We use $\theta$ for the angle between the $Z$- and the $z$-axis. We choose one of the two independent orthogonal directions to $Z$ to be denoted by $X$, located within the $xz$-plane and the other one, $Y$, coincides with the $y$-axis. We denote by $\varphi$ the polar angle in the $XY$-plane with respect to the $Y$-axis. The relation between the $XYZ$-coordinate and the $xyz$-coordinate is provided through the following transformation:

$$\begin{cases} z = Z\cos\theta - X\sin\theta, \\ x = Z\sin\theta + X\cos\theta, \\ y = Y. \end{cases} \quad (2)$$

If $p_\varphi$ is called the momentum component in the direction in the $XY$-plane specified by $\varphi$, then clearly

$$p_\varphi = p_Y \cos\varphi + p_X \sin\varphi. \quad (3)$$

We consider Eq. (3) as the differential treatment, then we square and average both sides of the equation instead. Because $\Delta p_Y \Delta p_X = 0$, the result is

$$<\Delta p_\varphi^2> = <\Delta p_Y^2>\cos^2\varphi + <\Delta p_X^2>\sin^2\varphi. \quad (4)$$

The definition of the jet quenching parameter $\hat{q}$ is the average momentum squared acquired by the quark through the medium at a unit distance [50–52]. So we obtain

$$\hat{q}_{\theta,\varphi} = \cos^2\varphi \hat{q}_{\theta,0} + \sin^2\varphi \hat{q}_{\theta,\frac{\pi}{2}}. \quad (5)$$

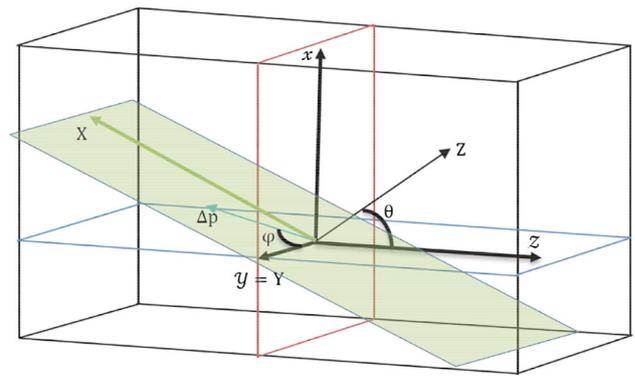

**Fig. 1** Schematic plot on relative orientation between the anisotropic direction $z$, the direction of motion of the quark, and the direction in which the momentum is measured $\triangle p$

## 3 Spatially anisotropic black brane with a chemical potential

In this section, we calculate the jet quenching parameter in the spatially anisotropic black brane background with a chemical potential [6,19,20]. We adopt the light-cone metric and use coordinates of $\sigma^\alpha(\tau, \sigma)$ to parameterize the worldsheet. Note that the action is invariant under a change. Then we calculate the jet quenching by the Nambu–Goto action. Afterwards we figure out the jet quenching parameter through the non-perturbative definition of $\hat{q}$. Finally, we examine the effect of the charge $Q$ and the anisotropic parameter $a$ by comparing results on varying $Q$ and $a$. In the beginning, we mainly consider the anisotropic action as follows:

$$I = \frac{1}{2\kappa^2} \int d^5x\sqrt{-g}\left(R + 12 - \frac{1}{2}(\partial\phi)^2 - \frac{1}{2}e^{2\alpha\phi}(\partial\chi)^2 - \frac{1}{4}F_{\mu\nu}F^{\mu\nu}\right) + S_{GH}, \quad (6)$$

where we have set $\kappa^2 = 4\pi^2/N_c^2$ and $S_{GH}$ is the Gibbons–Hawking boundary term. We obtain the metric as

$$ds_5^2 = \frac{L^2}{u^2}\left(-\mathcal{F}\mathcal{B}dt^2 + dx^2 + dy^2 + \mathcal{H}dz^2 + \frac{du^2}{\mathcal{F}}\right) + L^2 e^{\frac{1}{2}\phi}d\Omega_5^2, \quad (7)$$

$$A = A_t(u)dt, \chi = az, \phi = \phi(u), \quad (8)$$

where $L$ is the AdS radius and the event horizon is located at $u = u_H$. The functions $\mathcal{F}, \mathcal{B}$, and $\mathcal{H}$ can be found in [20]. The Hawking temperature is as follows:

$$T = -\frac{\mathcal{F}'(u_H)\sqrt{\mathcal{B}_H}}{4\pi}$$

$$= \sqrt{\mathcal{B}_H}\left[\frac{e^{-\frac{\phi_H}{2}}}{16\pi u_H}(16 + a^2 e^{7\frac{\phi_H}{2}}u_H^2) - \frac{e^{2\phi_H}Q^2 u_H^5}{24\pi}\right]. \quad (9)$$

According to the Bekenstein–Hawking entropy formula, the entropy density goes as

$$s = \frac{A_H}{4GV_3} = \frac{N_c^2 e^{-\frac{5}{4}\phi_H}}{2\pi u_H^3}, \qquad (10)$$

where $V_3$ is the volume of the black brane horizon.

### 3.1 Motion along the longitudinal direction

Using light-cone coordinates

$$z^\pm = \frac{t \pm z}{\sqrt{2}}, \qquad (11)$$

we can obtain the transformed metric as follows:

$$dS_5^2 = \frac{L^2}{u^2}\left[\frac{1}{2}(\mathcal{H}-\mathcal{FB})(dz^+)^2 + \frac{1}{2}(\mathcal{H}-\mathcal{FB})(dz^-)^2 \right.$$
$$\left. -(\mathcal{H}+\mathcal{FB})dz^+dz^- + dx^2 + dy^2 + \frac{du^2}{\mathcal{F}}\right]. \quad (12)$$

We consider a quark moving along $z^-$. The $xy$-plane is rotational symmetric and we can set $y = 0$ without loss of generality. Then we take the static gauge by identifying $(z^-, x) = (\tau, \sigma)$. We specify the string embedding through one function $u = u(x)$ subject to the boundary conditions $u(\pm l/2) = 0$. On this occasion the Nambu–Goto action takes the form

$$S = -\frac{1}{2\pi\alpha'}\int d\tau d\sigma \sqrt{-det g_{ind}}. \qquad (13)$$

Since we have

$$\det g_{ind} = \det g_{\alpha\beta} = \frac{L^4}{2u^4}(\mathcal{H}-\mathcal{FB})\left(1+\frac{u'^2}{\mathcal{F}}\right), \qquad (14)$$

and $L^- \gg L$, where $L^-$ is the long side of the Wilson loop, the action now reads

$$S = \frac{iL^2 L^-}{\pi\alpha'}\int_0^{l/2} dx \frac{1}{u^2}\sqrt{\frac{1}{2}(\mathcal{H}-\mathcal{FB})\left(1+\frac{u'^2}{\mathcal{F}}\right)}. \quad (15)$$

We use the Euler–Lagrange equation to deal with the integration. The $x$-independence of the Lagrangian results in a conserved quantity $\Pi_x$ and we just take the first-order term, which is

$$u'^2 = \frac{\mathcal{F}}{2\Pi_x^2 u^4}\left[(\mathcal{H}-\mathcal{FB}) - 2\Pi_x^2 u^4\right]. \qquad (16)$$

Since $u' = \partial_\sigma u$ and $\int_0^{l/2} d\sigma = l/2$, one finds

$$\frac{l}{2} = \sqrt{2}\Pi_x \int_0^{u_H} du \frac{u^2}{\sqrt{\mathcal{F}}\sqrt{(\mathcal{H}-\mathcal{FB}) - 2\Pi_x^2 u^4}}. \qquad (17)$$

Note that, as expected, $l \to 0$ as $\Pi_x \to 0$, so in this limit we obtain

$$l = 2\sqrt{2}\Pi_x \mathcal{I}_x + \mathcal{O}(\Pi_x^2), \qquad (18)$$

where

$$\mathcal{I}_x \equiv \int_0^{u_H} du \frac{u^2}{\sqrt{\mathcal{F}}\sqrt{\mathcal{H}-\mathcal{FB}}}. \qquad (19)$$

Substituting Eq. (16) into Eq. (15), one obtains

$$S = \frac{iL^2 L^-}{\sqrt{2}\pi\alpha'}\int_0^{u_H} \frac{du}{u^2} \frac{\mathcal{H}-\mathcal{FB}}{\sqrt{\mathcal{F}}\sqrt{(\mathcal{H}-\mathcal{FB}) - 2\Pi_x^2 u^4}}. \qquad (20)$$

The action diverges because the integration near $u = 0$. This can be seen by use of Maclaurin expansion in powers of $\Pi_x$, and we obtain the recast action

$$S = \frac{iL^2 L^-}{\sqrt{2}\pi\alpha'}\int_0^{u_H} du \frac{\sqrt{\mathcal{H}-\mathcal{BF}}}{u^2\sqrt{\mathcal{F}}} + \frac{iL^2 L^- l^2}{8\sqrt{2}\pi \mathcal{I}_x \alpha'} + \mathcal{O}(l^4). \qquad (21)$$

Substituting Eq. (18) into the above equation, and we have

$$S = \frac{iL^2 L^-}{\sqrt{2}\pi\alpha'}\left[\int_0^{u_H} \frac{du}{u^2}\frac{\sqrt{\mathcal{H}-\mathcal{FB}}}{\sqrt{\mathcal{F}}} + \mathcal{I}_x \Pi_x^2 + \mathcal{O}(\Pi_x^4)\right]. \qquad (22)$$

The jet quenching parameter can be obtained from the finite $l^2$-term which does not require any renormalization. It then follows that $\hat{q}$ is not sensitive to the presence of the anomaly. The relation between the action and the jet quenching parameter is given by [14,49]

$$e^{i2S} = <W^A(\mathcal{C})> \approx e^{-\frac{L^- l^2}{4\sqrt{2}}\hat{q}}, \qquad (23)$$

where $S$ signifies the finite part of the action.

Comparing the $l^2$ term in (21) with that in (23), we obtain

$$\hat{q}_z = \hat{q}_{0,\varphi} = \frac{L^2}{\pi \mathcal{I}_x \alpha'} = \frac{\sqrt{\lambda}}{\pi \mathcal{I}_x}, \qquad (24)$$

where $\sqrt{\lambda} = L^2/\alpha'$ is the coupling parameter.

For Eq. (24), we can solve the value of $\hat{q}_z$ by giving values of $Q$ and $a$. When $Q$ equals zero, the case will recover the situation of [15]. We have numerically determined $\hat{q}_z$ as a function of $a/T$ for various values of $Q$. In Fig. 2, we show the relation between the ratio of jet quenching parameter $\hat{q}_z$ to isotropic jet quenching parameter $\hat{q}_{iso}^{Q=0}$ and $a/T$ for $Q = 0, 0.5, 1, 3, 5$. We find that the jet quenching parameter is generally enhanced as the chemical potential increases. We can easily observe that the behavior of longitudinal jet quenching for large chemical potential is different from the small chemical potential case. For large chemical potential the jet quenching parameter is growing much faster than the small one at large $a/T$.

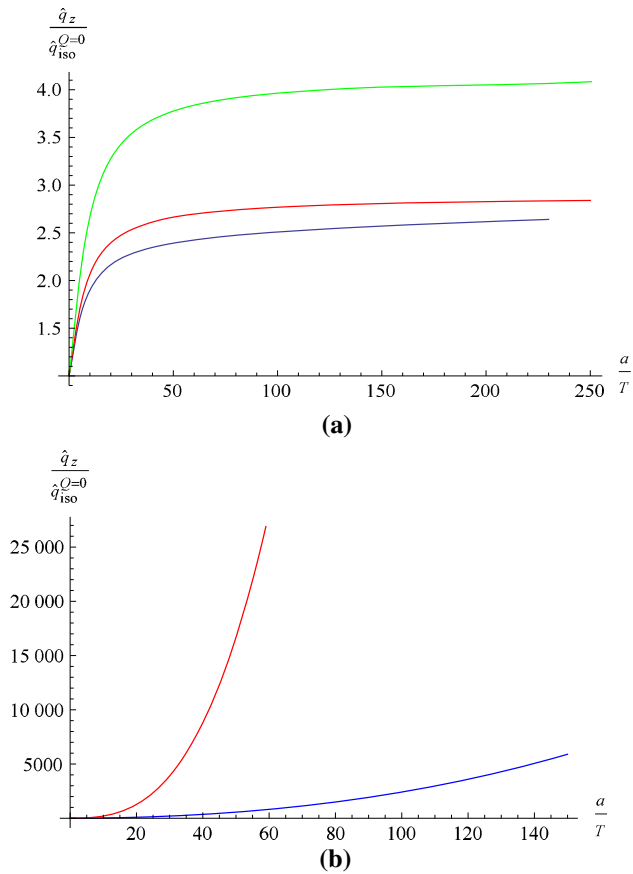

**Fig. 2** The function of the anisotropy of jet quenching parameter $\hat{q}_z$ for a quark moving along the longitudinal $z$-direction. **a** The ratio of $\hat{q}_z$ to $\hat{q}_{iso}^{Q=0}$ for $Q = 0$ (*blue*), $Q = 0.5$ (*red*) and $Q = 1$ (*green*). **b** Large $Q$ cases show results for $Q = 3$ and $Q = 5$ by a *blue line* and *red line*, respectively

### 3.2 Motion in the transverse plane

Then we discussed the situation where the string motion is in the transverse plane. Due to the rotational symmetry in $xy$-plane, we can simply choose the string moving in the $x$-direction. This situation corresponds to the case $\theta = \pi/2$. In the same way as in the previous example, it is convenient to compute with suitable light-cone coordinates,

$$x^{\pm} = \frac{t \pm x}{\sqrt{2}}. \tag{25}$$

The metric (7) now takes the form

$$dS_5^2 = \frac{L^2}{u^2} \left[ \frac{1}{2}(1 - \mathcal{F}\mathcal{B})(dx^+)^2 + \frac{1}{2}(1 - \mathcal{F}\mathcal{B})(dx^-)^2 \right.$$
$$\left. - (1 + \mathcal{F}\mathcal{B})dx^+ dx^- + dy^2 + \mathcal{H}dz^2 + \frac{du^2}{\mathcal{F}} \right]. \tag{26}$$

In this situation, we fix the static gauge by identifying $(x^-, u) = (\tau, \sigma)$. At the same time, $x^+ = constant$, and the string projection in the $xy$-plane can be specified as follows:

$$y \to \cos\varphi y(u), \quad z \to \sin\varphi z(u). \tag{27}$$

One obtains the following result by using the Nambu–Goto action (13):

$$S = \frac{iL^2}{\pi \alpha'} \int dx \int_0^{u_H} du$$
$$\times \frac{1}{u^2} \sqrt{\frac{1}{2}(1 - \mathcal{F}\mathcal{B})\left(\frac{1}{\mathcal{F}} + y'^2 \cos^2\varphi + \mathcal{H}z'^2 \sin^2\varphi\right)}. \tag{28}$$

Similarly, since the Lagrangian does not depend on $y, z$ explicitly, we find that

$$y' = \frac{\sqrt{2\mathcal{H}}u^2 \Pi_y}{\sqrt{\mathcal{F}}\sqrt{\mathcal{H}(1 - \mathcal{F}\mathcal{B}) - 2u^4\left(\mathcal{H}\Pi_y^2 \cos^2\varphi + \Pi_z^2 \sin^2\varphi\right)}}, \tag{29}$$

$$z' = \frac{\sqrt{2}u^2 \Pi_z}{\sqrt{\mathcal{H}\mathcal{F}}\sqrt{\mathcal{H}(1 - \mathcal{F}\mathcal{B}) - 2u^4\left(\mathcal{H}\Pi_y^2 \cos^2\varphi + \Pi_z^2 \sin^2\varphi\right)}}, \tag{30}$$

where $\Pi_y$ and $\Pi_z$ are conserved quantities. Because of $y' = \partial_\sigma y$, $z' = \partial_\sigma z$, and $\int_0^{l/2} d\sigma = l/2$, following a similar procedure to the previous section, we obtain

$$\frac{l}{2} = \int_0^{u_H} du$$
$$\times \frac{\sqrt{2\mathcal{H}}u^2 \Pi_y}{\sqrt{\mathcal{F}}\sqrt{\mathcal{H}(1 - \mathcal{F}\mathcal{B}) - 2u^4\left(\mathcal{H}\Pi_y^2 \cos^2\varphi + \Pi_z^2 \sin^2\varphi\right)}}, \tag{31}$$

$$\frac{l}{2} = \int_0^{u_H} du$$
$$\times \frac{\sqrt{2}u^2 \Pi_z}{\sqrt{\mathcal{H}\mathcal{F}}\sqrt{\mathcal{H}(1-\mathcal{F}\mathcal{B}) - 2u^4\left(\mathcal{H}\Pi_y^2 \cos^2\varphi + \Pi_z^2 \sin^2\varphi\right)}}. \tag{32}$$

Expanding the above two equations in the limit $\Pi_y \to 0$ and $\Pi_z \to 0$, respectively,

$$l = 2\sqrt{2}\Pi_y \mathcal{I}_{xy} + \mathcal{O}(\Pi^2), \tag{33}$$

$$l = 2\sqrt{2}\Pi_z \mathcal{I}_{xz} + \mathcal{O}(\Pi^2), \tag{34}$$

with

$$\mathcal{I}_{xy} \equiv \int_0^{u_H} du \frac{u^2}{\sqrt{\mathcal{F}}(1 - \mathcal{F}\mathcal{B})}, \tag{35}$$

$$\mathcal{I}_{xz} \equiv \int_0^{u_H} du \frac{u^2}{\mathcal{H}\sqrt{\mathcal{F}}(1 - \mathcal{F}\mathcal{B})}, \tag{36}$$

which are convergent integrals. Substituting Eq. (29) and Eq. (30) into Eq. (28), we can obtain

$$S = \frac{iL^2 L^-}{\sqrt{2}\pi\alpha'} \int_0^{u_h} du \times \frac{\sqrt{\mathcal{H}}(1-\mathcal{FB})}{u^2\sqrt{\mathcal{F}}(\mathcal{H}(1-\mathcal{FB}) - 2u^4(\mathcal{H}\Pi_y^2\cos^2\varphi + \Pi_z^2\sin^2\varphi))^{\frac{1}{2}}}. \quad (37)$$

Again, we can expand the action in powers of $\Pi_y$ and $\Pi_z$, and obtain the result to the second order:

$$S = \frac{iL^2 L^-}{\sqrt{2}\pi\alpha'} \int_0^{u_H} du \frac{\sqrt{1-\mathcal{FB}}}{u^2\sqrt{\mathcal{F}}} + \frac{iL^2 L^-}{\sqrt{2}\pi\alpha'} \int_0^{u_H} du \frac{u^2\mathcal{H}\cos^2\varphi\Pi_y^2 + u^2\sin^2\varphi\Pi_z^2}{\mathcal{H}\sqrt{\mathcal{F}}\sqrt{1-\mathcal{FB}}}. \quad (38)$$

Similarly, we only take out the second item. Because $\Pi_y = l/2\sqrt{2}\mathcal{I}_{xy}$ and $\Pi_z = l/2\sqrt{2}\mathcal{I}_{xz}$, the action becomes

$$S = \frac{iL^2 L^- l^2}{8\sqrt{2}\pi\alpha'} \left( \frac{\cos^2\varphi}{\mathcal{I}_{xy}} + \frac{\sin^2\varphi}{\mathcal{I}_{xz}} \right). \quad (39)$$

Following the prescription Eq. (23), the transverse jet quenching parameter is given by

$$\hat{q}_{xy} = \hat{q}_{\frac{\pi}{2},0} = \frac{L^2}{\pi\alpha'\mathcal{I}_{xy}}, \hat{q}_{xz} = \hat{q}_{\frac{\pi}{2},\frac{\pi}{2}} = \frac{L^2}{\pi\alpha'\mathcal{I}_{xz}}. \quad (40)$$

Hence

$$\hat{q}_{\frac{\pi}{2},\varphi} = \hat{q}_{xy}\cos^2\varphi + \hat{q}_{xz}\sin^2\varphi. \quad (41)$$

In Fig. 3, we find that, for a small chemical potential, there exists a peak in $\hat{q}_{xy}$ at low anisotropy $a/T$, then $\hat{q}_{xy}$ decreases and becomes lower than the isotropic case at large anisotropy $a/T$. But when the chemical potential is larger, $\hat{q}_{xy}$ behaves different from the small chemical potential case. It always increases and grows fast as the anisotropy $a/T$ increases.

In Fig. 4, we show the jet quenching parameter $\hat{q}_{xz}$ for a quark moving in the transverse $xy$ plane. In contrast to $\hat{q}_{xy}$, $\hat{q}_{xz}$ always increases whether the chemical potential is large or not. In addition, we have plotted $\hat{q}_{\frac{\pi}{2},\varphi}$ to $\hat{q}_{iso}^{Q=0}$ for various values of $\varphi$ in Fig. 5.

The drag force calculated in [16] shows that the jet quenching increases with the increasing charge $Q$. Comparing with the jet quenching computed in [6], we find that the jet quenching is also increasing as the charge density $Q$ increases.

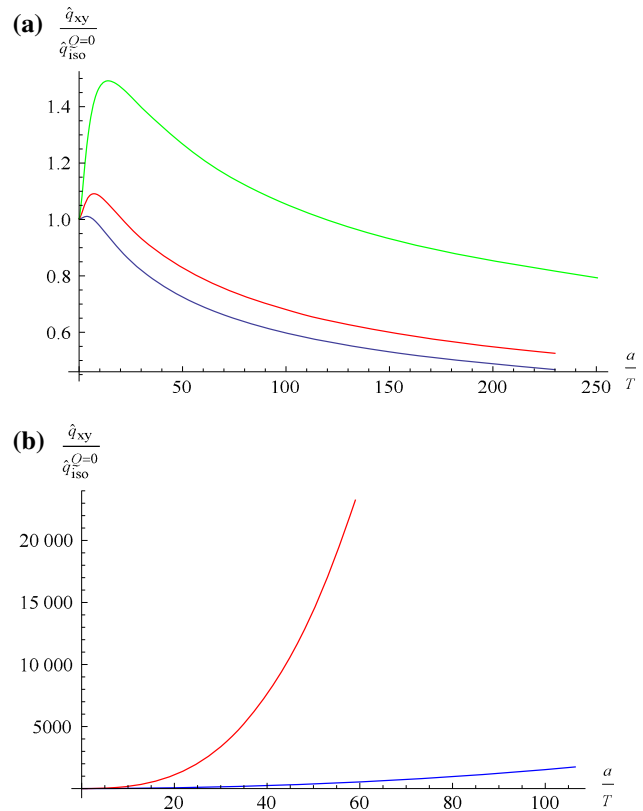

**Fig. 3** Functions of the anisotropic jet quenching parameter $\hat{q}_{xy}$ for a quark moving along the vertical direction of the $XY$-plane. The $Y$-axis is the ratio of the jet quenching parameter $\hat{q}_{xy}$ to $\hat{q}_{iso}^{Q=0}$, the $X$-axis is the ratio of parameter $a$ to temperature $T$. **a** *Lines* show results for $Q = 0$ (*blue*), $Q = 0.5$ (*red*) and $Q = 1$ (*green*). **b** *Blue line* and *red line* describe results for $Q = 3$ and $Q = 5$ respectively

## 4 Anisotropic black brane in the IR region

In this section, we will consider the jet quenching in the Einstein–dilaton–axion model discussed in [53]. The Einstein–dilaton–axion action is given by

$$S = \frac{1}{2\kappa^2} \int d^5x \sqrt{-g} \left( R + 12\Lambda - \frac{1}{2}(\partial\phi)^2 - \frac{1}{2}e^{2\alpha\phi}(\partial\chi)^2 \right). \quad (42)$$

Here $2\kappa^2 = 16\pi G$ is the gravitational coupling and $G$ is the Newton constant in 5 dimensions. The parameter $\alpha$ enters in the dilaton dependence of axion kinetic term. Earlier work in [39] considered the case with $\alpha = 1$, and the case $\alpha = -1$ has $SL(2, R)$ invariance. It is easy to see there exists a near-extremal solution whose near horizon limit at small temperature $T \ll \rho$ is given by

$$ds^2 = \frac{R^2}{u^2}\left[ -\mathcal{F}dt^2 + \frac{du^2}{\mathcal{F}} + dx^2 + dy^2 + \mathcal{H}dz^2 \right], \quad (43)$$

$$\mathcal{F} = 1 - \left(\frac{u}{u_h}\right)^p, \quad \mathcal{H} = \rho^2 u^{\frac{2}{1+2\alpha^2}}, \quad R^2 = \frac{3+8\alpha^2}{4+8\alpha^2}, \quad (44)$$

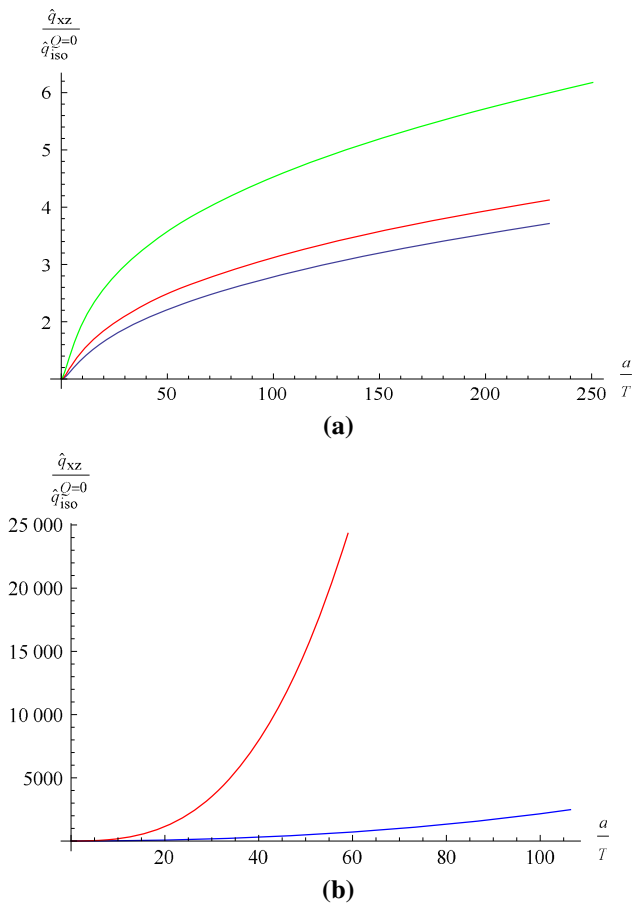

**Fig. 4** Functions of the anisotropic jet quenching parameter $\hat{q}_{xz}$ for a quark moving along the parallel direction of the $XY$-plane. The *vertical* coordinate is the ratio of the jet quenching parameter $\hat{q}_{xz}$ to $\hat{q}_{iso}^{Q=0}$, the *horizontal* coordinate is the ratio of parameter $a$ to temperature $T$. **a** *Lines* show results for $Q = 0$ (*blue*), $Q = 0.5$ (*red*) and $Q = 1$ (*green*). **b** *Blue line* and *red line* describe results for $Q = 3$ and $Q = 5$, respectively

$$p = \frac{3+8\alpha^2}{1+2\alpha^2}, \quad \chi = c_1 \rho z, \tag{45}$$

$$\phi = -\frac{2\alpha}{1+2\alpha^2}\log(u), \quad c_1 = \frac{\sqrt{2(3+8\alpha^2)}}{(1+2\alpha^2)}. \tag{46}$$

The metric is a special case of the metric (1), which corresponds to $z = 1/2\alpha^2 + 1$. The black hole horizon is at $u = u_h$ and the boundary is at $u = 0$. The Hawking temperature is given by

$$T = \frac{p^2}{16\pi u_h}. \tag{47}$$

This solution breaks rotational invariance along the $z$-direction due to the linearly varying axion, and $\rho$ is the mass scale which characterizes the breaking of anisotropy.

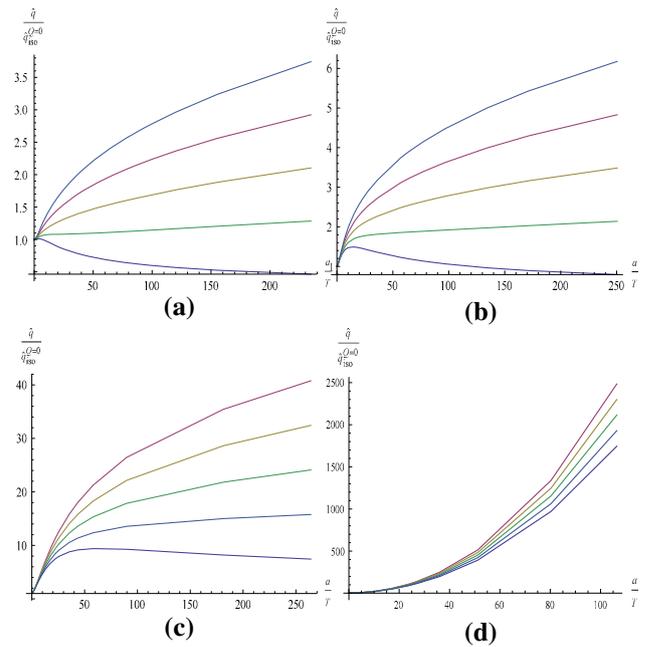

**Fig. 5 a** The transverse jet quenching parameter $\hat{q}_{\pi/2,\varphi}$ for $Q = 0$, from *top* to *bottom*, with $\varphi = \pi/2, \pi/3, \pi/4, \pi/6, 0$. **b** The transverse jet quenching parameter $\hat{q}_{\pi/2,\varphi}$ for $Q = 1$, from *top* to *bottom*, corresponds to $\varphi = \pi/2, \pi/3, \pi/4, \pi/6, 0$, respectively. **c** The transverse jet quenching parameter $\hat{q}_{\pi/2,\varphi}$ for $Q = 2$, from *top* to *bottom*, with $\varphi = \pi/2, \pi/3, \pi/4, \pi/6, 0$. **d** The transverse jet quenching parameter $\hat{q}_{\pi/2,\varphi}$ for $Q = 3$, from *top* to *bottom*, with $\varphi = \pi/2, \pi/3, \pi/4, \pi/6, 0$

### 4.1 Motion along the longitudinal direction

To compute the jet quenching parameter along the longitudinal direction, we use the light-cone coordinate so that

$$z^{\pm} = \frac{t \pm z}{\sqrt{2}}, \tag{48}$$

and we can recast the metric as

$$dS^2 = \frac{R^2}{u^2}\left[\frac{1}{2}(\mathcal{H}-\mathcal{F})(dz^+)^2 + \frac{1}{2}(\mathcal{H}-\mathcal{F})(dz^-)^2 \right.$$
$$\left. -(\mathcal{H}+\mathcal{F})dz^+dz^- + dx^2 + dy^2 + \frac{du^2}{\mathcal{F}}\right]. \tag{49}$$

Now we consider a quark moving along the $z^-$-direction. The $xy$-plane is symmetric and we can set $y = 0$ without loss of generality. Then we fix the static gauge by ascertaining $(z^-, x) = (\tau, \sigma)$, and assign the string embedding through one function $u = u(x)$, which is subjected to the boundary condition $u(\pm l/2) = 0$. Note that the Nambu–Goto action is given by

$$S = -\frac{1}{2\pi\alpha'}\int d\tau d\sigma \sqrt{-\det g_{ind}}. \tag{50}$$

Because of

$$\det g_{\alpha\beta} = \frac{R^4}{2u^4}(\mathcal{H}-\mathcal{F})\left(1+\frac{u'^2}{\mathcal{F}}\right), \quad L^- \gg L, \tag{51}$$

the action takes the form

$$S = \frac{iR^2 L^-}{\pi \alpha'} \int_0^{l/2} dx \frac{1}{u^2} \sqrt{\frac{1}{2}(\mathcal{H} - \mathcal{F})\left(1 + \frac{u'^2}{\mathcal{F}}\right)}. \quad (52)$$

Following the same procedure, noting that the Lagrangian does not depend on $x$ explicitly leads to a conserved quantity $\mathcal{E}_x$, and we take the first-order term, that is to say:

$$u'^2 = \frac{\mathcal{F}}{2\mathcal{E}_x^2 u^4}[(\mathcal{H} - \mathcal{F}) - 2\mathcal{E}_x^2 u^4]. \quad (53)$$

Since $u' = \partial_\sigma u$ and $\int_0^{l/2} d\sigma = l/2$, one finds

$$\frac{l}{2} = \sqrt{2}\mathcal{E}_x \int_0^{u_h} du \frac{u^2}{\sqrt{\mathcal{F}}\sqrt{(\mathcal{H} - \mathcal{F}) - 2\mathcal{E}_x^2 u^4}}. \quad (54)$$

Note that $l \to 0$ as $\mathcal{E}_x \to 0$, so in this limit we can obtain

$$l = 2\sqrt{2}\mathcal{E}_x \mathcal{K}_x + \mathcal{O}(\mathcal{E}_x^2), \quad (55)$$

where

$$\mathcal{K}_x \equiv \int_0^{u_h} du \frac{u^2}{\sqrt{\mathcal{F}}\sqrt{\mathcal{H} - \mathcal{F}}}. \quad (56)$$

We substitute Eq. (53) into Eq. (52), then we obtain the equation as follows:

$$S = \frac{iR^2 L^-}{\sqrt{2}\pi \alpha'} \int_0^{u_h} \frac{du}{u^2} \frac{\mathcal{H} - \mathcal{F}}{\sqrt{\mathcal{F}}\sqrt{(\mathcal{H} - \mathcal{F}) - 2\mathcal{E}_x^2 u^4}}. \quad (57)$$

We expand the integrand and use Eqs. (55) and (56) to get

$$S = \frac{iR^2 L^-}{\sqrt{2}\pi \alpha'} \int_0^{u_h} du \frac{\sqrt{\mathcal{H} - \mathcal{F}}}{u^2 \sqrt{\mathcal{F}}} + \frac{iR^2 L^- l^2}{8\sqrt{2}\pi \mathcal{K}_x \alpha'} + \mathcal{O}(l^4). \quad (58)$$

As we have

$$e^{i2S} = <W^A(\mathcal{C})> \approx exp[-\frac{L^- l^2}{4\sqrt{2}}\hat{q}], \quad (59)$$

this leads to

$$\hat{q}_z = \hat{q}_{0,\varphi} = \frac{L^2}{\pi \mathcal{K}_x \alpha'} = \frac{\sqrt{\lambda}}{\pi \mathcal{K}_x}, \quad (60)$$

where $\sqrt{\lambda} = R^2/\alpha'$.

Figure 6 shows the jet quenching parameter $\hat{q}_z$ as a function of $\alpha$. It is found that as $\alpha$ increases, the ratio increases and approaches a fixed value at large $\alpha$.

### 4.2 Motion in the transverse plane

Next we discuss the situation where the motion is in the transverse plane. We set the rotational symmetry in the xy-plane and choose the motion in x-direction. That is $\theta = \pi/2$ in Fig. 1. In the y-direction and z-direction, there is no rotational symmetry, so in this situation the result will be dependent of $\varphi$. In the same way as in the previous example, we consider the light-cone coordinates,

$$x^\pm = \frac{t \pm x}{\sqrt{2}}. \quad (61)$$

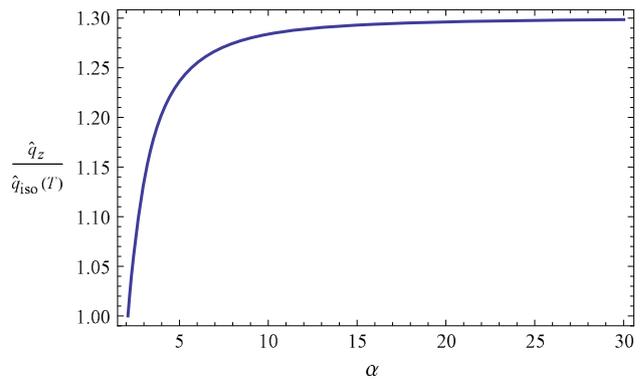

**Fig. 6** The *vertical* coordinate is the ratio of the jet quenching parameter of anisotropic black brane to the isotropic one. The *horizontal* coordinate is $\alpha$, which can be used to describe the change of the geometry near horizon

The metric (7) becomes

$$dS^2 = \frac{R^2}{u^2}\left[\frac{1}{2}(1-\mathcal{F})(dx^+)^2 + \frac{1}{2}(1-\mathcal{F})(dx^-)^2 \right.$$
$$\left. -(1+\mathcal{F})dx^+ dx^- + dy^2 + \mathcal{H}dz^2 + \frac{du^2}{\mathcal{F}}\right]. \quad (62)$$

In this situation, we fix the static gauge by identifying $(x^-, u) = (\tau, \sigma)$. At the same time, $x^+ = const.$, and we specify the string projection in the xy-plane as follows:

$$y \to \cos\varphi y(u), \quad z \to \sin\varphi z(u). \quad (63)$$

One can get the following result by using (50)

$$S = \frac{iR^2}{\pi \alpha'} \int dx \int_0^{u_h} du$$
$$\times \frac{1}{u^2}\sqrt{\frac{1}{2}(1-\mathcal{F})\left(\frac{1}{\mathcal{F}} + y'^2 \cos^2\varphi + \mathcal{H}z'^2 \sin^2\varphi\right)}. \quad (64)$$

The fact that the Lagrangian does not have explicit an $y, z$ dependence leads to

$$y' = \frac{\sqrt{2\mathcal{H}}u^2 \mathcal{E}_y}{\sqrt{\mathcal{F}}\sqrt{\mathcal{H}(1-\mathcal{F}) - 2u^4(\mathcal{H}\mathcal{E}_y^2 \cos^2\varphi + \mathcal{E}_z^2 \sin^2\varphi)}}, \quad (65)$$

and

$$z' = \frac{\sqrt{2}u^2 \mathcal{E}_z}{\sqrt{\mathcal{H}\mathcal{F}}\sqrt{\mathcal{H}(1-\mathcal{F}) - 2u^4(\mathcal{H}\mathcal{E}_y^2 \cos^2\varphi + \mathcal{E}_z^2 \sin^2\varphi)}}, \quad (66)$$

where $\mathcal{E}_y$ and $\mathcal{E}_z$ are conserved quantities. Because $y' = dy/du$ and $z' = dz/du$, $\int_0^{l/2} dy = l/2$ and $\int_0^{l/2} dz = 1/2$, one has

$$\frac{l}{2} = \sqrt{2}\mathcal{E}_y \int_0^{u_h} du$$
$$\times \frac{\sqrt{\mathcal{H}}u^2}{\sqrt{\mathcal{F}}\sqrt{\mathcal{H}(1-\mathcal{F}) - 2u^4(\mathcal{H}\mathcal{E}_y^2\cos^2\varphi + \mathcal{E}_z^2\sin^2\varphi)}}, \quad (67)$$

$$\frac{l}{2} = \sqrt{2}\mathcal{E}_z \int_0^{u_h} du$$
$$\times \frac{u^2}{\sqrt{\mathcal{H}\mathcal{F}}\sqrt{\mathcal{H}(1-\mathcal{F}) - 2u^4(\mathcal{H}\mathcal{E}_y^2\cos^2\varphi + \mathcal{E}_z^2\sin^2\varphi)}}. \quad (68)$$

We expand the above two equations in the limit $\mathcal{E}_y \to 0$ and $\mathcal{E}_z \to 0$, respectively,

$$l = 2\sqrt{2}\mathcal{E}_y\mathcal{K}_{xy} + \mathcal{O}(\mathcal{E}^2), \quad (69)$$

$$l = 2\sqrt{2}\mathcal{E}_z\mathcal{K}_{xz} + \mathcal{O}(\mathcal{E}^2), \quad (70)$$

with

$$\mathcal{K}_{xy} \equiv \int_0^{u_h} du \frac{u^2}{\sqrt{\mathcal{F}(1-\mathcal{F})}}, \quad (71)$$

$$\mathcal{K}_{xz} \equiv \int_0^{u_h} du \frac{u^2}{\mathcal{H}\sqrt{\mathcal{F}(1-\mathcal{F})}}. \quad (72)$$

Take Eqs. (65) and (66) into Eq. (64),

$$S = \frac{iR^2L^-}{\sqrt{2}\mathcal{E}\alpha'} \int_0^{u_h} du \frac{\sqrt{\mathcal{H}(1-\mathcal{F}\mathcal{H})}^2}{u}$$
$$\times \left[\frac{\mathcal{F}}{\mathcal{H}(1-\mathcal{F}) - 2u^4(\mathcal{E}_y^2\cos^2\varphi + \mathcal{E}_z^2\sin^2\varphi)}\right]^{\frac{1}{2}}. \quad (73)$$

We expand the integrand and rearrange the above equation, and we obtain

$$S = \frac{iR^2L^-}{\sqrt{2}\pi\alpha'} \int_0^{u_h} du \frac{\sqrt{1-\mathcal{F}}}{u^2\sqrt{\mathcal{F}}}$$
$$+ \frac{iR^2L^-}{\sqrt{2}\pi\alpha'} \int_0^{u_h} du \frac{u^2\mathcal{H}\cos^2\varphi\mathcal{E}_y^2 + u^2\sin^2\varphi\mathcal{E}_z^2}{\mathcal{H}\sqrt{\mathcal{F}}\sqrt{1-\mathcal{F}}}. \quad (74)$$

We only take out the second item, then

$$S = \frac{iR^2L^-}{\sqrt{2}\pi\alpha'}(\cos^2\varphi\mathcal{E}_y^2\mathcal{K}_{xy} + \sin^2\varphi\mathcal{E}_z^2\mathcal{K}_{xz}), \quad (75)$$

and because $\mathcal{E}_y = l/2\sqrt{2}\mathcal{K}_{xy}$ and $\mathcal{E}_z = l/2\sqrt{2}\mathcal{K}_{xz}$, this yields

$$S = \frac{iR^2L^-l^2}{8\sqrt{2}\pi\alpha'}\left(\frac{\cos^2\varphi}{\mathcal{K}_{xy}} + \frac{\sin^2\varphi}{\mathcal{K}_{xz}}\right) \quad (76)$$

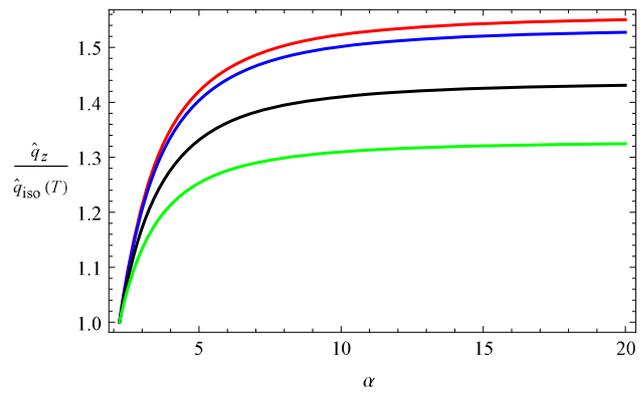

**Fig. 7** $\hat{q}_z$ as a function of $\alpha$ with different $\varphi$. From *top* to *bottom*, the *curves* represent $\varphi = \pi/2, \pi/3, \pi/6, 0$ relative to the $y$-axis

and

$$\hat{q}_\perp = \hat{q}_{\frac{\pi}{2},0} = \frac{R^2}{\pi\alpha'\mathcal{K}_{xy}}, \quad \hat{q}_L = \hat{q}_{\frac{\pi}{2},\frac{\pi}{2}} = \frac{R^2}{\pi\alpha'\mathcal{K}_{xz}}. \quad (77)$$

As a result,

$$\hat{q}_{\frac{\pi}{2},\varphi} = \hat{q}_\perp \cos^2\varphi + \hat{q}_L \sin^2\varphi. \quad (78)$$

From Fig. 7, we find the same situation as for the longitudinal direction, values of the ratio rising with the increasing of $\alpha$. When it reaches one value, it stops rising, in spite of the growth of $\alpha$. We know the values of the ratio are reduced as the angle $\varphi$ decreases.

We find that our results are consistent with results in [14,15]. We know the jet quenching parameter as the temperature increases from [14,15]. In our case, the parameter $\alpha^2$ is proportional to the temperature so that our result is consistent with [14,15].

## 5 Summary

The jet quenching parameter $\hat{q}$ describes the momentum broadening of a highly relativistic parton moving through a non-Abelian plasma. We have considered two cases, one is a R-charged version of the anisotropic black brane [19,20] and the other is an anisotropic black brane in the IR region [39,40]. Both are in the coordinate system, where there exists rotational symmetry in the $x$, $y$-directions but not in the $z$-direction. In the context of heavy ion collisions the former $x$, $y$ symmetric directions and the latter $z$ asymmetric directions would correspond to the transverse plane and beam direction, respectively. There are some factors that affect the jet quenching parameter as follows: (1) the relative orientation between these directions, (2) the movement direction of the parton, (3) the direction of the measured momentum broadening, and (4) the charge density $Q$ or, equivalently, the chemical potential. We mainly discuss the quark moving in the longitudinal direction and the transverse plane, namely we study the cases $\theta = 0$ and $\theta = \pi/2$.

In Sect. 3, we compute the jet quenching parameter in the background of a charged anisotropic black brane. We find that for small enough anisotropy the jet quenching parameter is always larger than that in an isotropic plasma at the same temperature, regardless of the directions of motion and momentum broadening. As the anisotropy increase, the momentum broadening for the moving quark will depend on the direction of the momentum broadening. In other words, at small $a/T$ the momentum broadening is larger for quarks propagating along the beam axis z, while it is larger for quarks propagating in the transverse plane for large $a/T$ (unless the momentum broadening is measured very close to the orthogonal direction within the transverse plane). On the other hand, the jet quenching parameter is always enhanced by the charge density $Q$, and the jet quenching parameter has different behaviors for small charge density and large charge density. When the charge density is small, the jet quenching parameter in the longitudinal direction will increase as $a/T$ increases and possibly approaches some constant at large $a/T$. But when the charge density is large, the jet quenching parameter in the longitudinal direction will increase fast at large $a/T$. As the quark is moving in the transverse plane, the jet quenching parameter has a similar behavior to the longitudinal case, i.e., it is always enhanced by the charge density and has different behaviors for small and large charge density. However, for the transverse case, the jet quenching parameter will depend on the direction of momentum broadening.

In Sect. 4, we compute the jet quenching parameter in the background in the IR region of a spatially anisotropic black hole in Einstein–dilaton–axion theory. For this case, we also conclude that the jet quenching parameter increases as the anisotropy $\alpha$ and tends to saturate at large anisotropy $\alpha$.

According to the above discussion, we can conclude that: (1) The jet quenching parameter of the anisotropic systems increases as the anisotropic parameter $a$ or $\alpha$ goes up. (2) The jet quenching parameter of the anisotropic systems also increases as the charge $Q$ goes up. (3) The jet quenching parameter increases with the increasing angle $\phi$. (4) When the $Q$ is small ($Q \leq 1$), the jet quenching parameter increases with increasing $a$ or $\alpha$ until to some fixed value. It illustrates that $a$ and $\alpha$ only impact the parameter at small value.

**Acknowledgements** LYW wish to thank Xian-Hui Ge for his support and helpful discussions during this work. The authors were partly supported by NSFC, China (No. 11375110) and the Grant from Shanghai Key Laboratory of High Temperature Superconductors with No. 14DZ2260700. SYW was partially supported by the Ministry of Science and Technology (Grant No. MOST 104-2811-M-009-068) and National Center for Theoretical Sciences in Taiwan.

**References**

1. J.M. Maldacena, The large $\mathcal{N}$ limit of superconformal field theories and supergravity. Adv. Theor. Math. Phys. **2**, 231 (1998) [Int. J. Theor. Phys. **38** 1113 (1999)]
2. S.S. Gubser, I.R. Klebanov, A.M. Polyakov, Gauge theory correlators from non-critical string theory. Phys. Lett. B **428**, 105 (1998)
3. E. Witten, Anti-de Sitter space and holography. Adv. Theor. Math. Phys. **2**, 253 (1998)
4. C.P. Herzog, A. Karch, P. Kovtun, C. Kozcaz, L.G. Yaffe, Energy loss of a heavy quark moving through $\mathcal{N} = \triangle$ supersymmetric Yang-Mills plasma. JHEP **0607**, 013 (2006). arXiv:hep-ph/0605158
5. S.S. Gubser, Drag force in AdS/CFT. Phys. Rev. D **74**, 126005 (2006)
6. L. Cheng, X.-H. Ge, W. Shang-Yu, Drag force of anisotropic plasma at finite U(1) chemical potential. Eur. Phys. J. C **76**, 256 (2016)
7. S. Chakraborty, N. Haque, Drag force in strongly coupled, anisotropic plasma at finite chemical potential. JHEP **1412**, 175 (2014). arXiv:1410.7040 [hep-th]
8. R.-G. Cai, S. Chakrabortty, S. He, L. Li, Some aspects of QGP phase in a hQCD model. JHEP **02**, 068 (2013). arXiv:1209.4512 [hep-th]
9. S. Caron-Huot, P. Kovtun, G.D. Moore, A. Starinets, L.G. Yaffe, Photon and dilepton production in supersymmetric Yang-Mills plasma. JHEP **0612**, 015 (2006)
10. L. Patino, D. Trancanelli, Thermal photon production in a strongly coupled anisotropic plasma. JHEP **1302**, 154 (2013)
11. S.Y. Wu, D.L. Yang, Holographic photon production with magnetic field in anisotropic plasmas. JHEP **1308**, 032 (2013)
12. B. Muller, S.Y. Wu, D.L. Yang, Elliptic flow from thermal photons with magnetic field in holography. Phys. Rev. D **89**, 026013 (2014)
13. S.-J. Sin, I. Zahed, Holography of radiation and jet quenching. Phys. Lett. B **608**, 265 (2005)
14. H. Liu, K. Rajagopal, U.A. Wiedemann, Calculating the jet quenching parameter from AdS/CFT. Phys. Rev. Lett. **97**, 182301 (2006)
15. M. Chernicoff, D. Fernandez, D. Mateos, D. Trancanelli, Jet quenching in a strongly coupled anisotropic plasma. JHEP **1210**, 041 (2012)
16. F.-L. Lin, T. Matsuo, Jet quenching parameter in medium with chemical potential from AdS/CFT. Phys. Lett. B **641**, 45 (2006)
17. K.B. Fadafan, H. Soltanpanahi, Energy loss in a strongly coupled anisotropic plasma. JHEP **10**, 085 (2012). arXiv:1206.2271 [hep-th]
18. S. Chakrabortty, T. K. Dey, Back reaction effects on the dynamics of heavy probes in heavy quark cloud. arXiv:1602.04761 [hep-th]
19. L. Cheng, X.-H. Ge, S.-J. Sin, Anisotropic plasma at finite U(1) chemical potential. JHEP **1407**, 083 (2014)
20. L. Cheng, X.-H. Ge, S.-J. Sin, Anisotropic plasma with a chemical potential and scheme -independent instabilities. Phys. Lett. B **734**, 116 (2014)
21. X.-H. Ge, Notes on shear viscosity bound violation in anisotropic models. Chin. Phys. Mech. Astron. **59**, 630401 (2016)
22. X.-H. Ge, Y. Ling, C. Niu, S.-J. Sin, Thermoelectric conductivities, shear viscosity, and stability in an anisotropic linear axion model. Phys. Rev. D **92**, 106005 (2015)
23. D. Giataganas, Probing strongly coupled anisotropic plasma. JHEP **1207**, 031 (2012)
24. W. Florkowski, Anisotropic fluid dynamics in the early stage of relativistic heavy-ion collisions. Phys. Lett. B **668**, 32 (2008)
25. W. Florkowski, R. Ryblewski, Dynamics of anisotropic plasma at the early stages of relativistic heavy-ion collisions. Acta Phys. Polon. B **40**, 2843 (2009)

26. R. Ryblewski, W. Florkowski, Early anisotropic hydrodynamics and the RHIC early-thermalization and HBT puzzles. Phys. Rev. C **82**, 024903 (2010)
27. W. Florkowski, R. Ryblewski, Highly-anisotropic and strongly-dissipative hydrodynamics for early stages of relativistic heavy-ion collisions. Phys. Rev. C **83**, 034907 (2011)
28. M. Martinez, M. Strickland, Dissipative dynamics of highly anisotropic systems. Nucl. Phys. A **848**, 183 (2010)
29. R. Ryblewski, W. Florkowski, Non-boost-invariant motion of dissipative and highly anisotropic fluid. J. Phys. G **38**, 015104 (2011)
30. M. Martinez, M. Strickland, Non-boost-invariant anisotropic dynamics. Nucl. Phys. A **856**, 68 (2011)
31. R. Ryblewski, W. Florkowski, Highly anisotropic hydrodynamics-discussion of the model assumptions and forms of the initial conditions. Acta Phys. Polon. B **42**, 115 (2011)
32. R. Ryblewski, W. Florkowski, Highly anisotropic and strongly-dissipative hydrodynamics with transverse expansion. Eur. Phys. J. C **71**, 1761 (2011)
33. I. Gahramanov, T. Kalaydzhyan, I. Kirsch, Anisotropic hydrodynamics, holography and the chiral magnetic effect. Phys. Rev. D **85**, 126013 (2012)
34. J.M. Maldacena, The large $\mathcal{N}$ limit of superconformal field theories and supergravity. Adv. Theor. Math. Phys. **2**, 231 (1998) [Int. J. Theor. Phys. **38** 1113 (1999)]
35. STAR collaboration, J. Adams et al., Experimental and theoretical challenges in the search for the quark gluon plasma:the STAR cllaboration's critical assessment of the evidence from RHIC collisions. Nucl. Phys. A **757**, 102 (2005)
36. PHENIX collaboration, K. Adcox et al., Formation of dense partonic matter in relativistic nucleus–nucleus collisions at RHIC:experimental evaluation by the PHENIX collaboration. Nucl. Phys. A **757**, 184 (2005)
37. E. Shuryak, Why does the quark gluon plasma at RHIC behave as a nearly ideal fluid? Progr. Part Nucl. Phys. **53**, 273 (2004)
38. E.V. Shuryak, What RHIC experiments and theory tell us about properties of quark-gluon plasma? Nucl. Phys. A **750**, 64 (2005)
39. D. Mateos, D. Trancanelli, Thermodynamics and instabilities of a strongly coupled anisotropic plasma. JHEP **1107**, 054 (2011)
40. D. Mateos, D. Trancanelli, The anisotropic $\mathcal{N} = 4$ super Yang–Mills plasma and its instabilities. Phys. Rev. Lett. **107**, 101601 (2011)
41. T. Azeyanagi, W. Li, T. Takayanagi, On string theory duals of Lifshitz-like fixed points. JHEP **0906**, 084 (2009)
42. S. Kachru, X. Liu, M. Mulligan, Gravity duals of Lifshitz-like fixed points. Phys. Rev. D **78**, 106005 (2008)
43. J.R. Sun, S.Y. Wu, H.Q. Zhang, Novel features of the transport coefficients in Lifshitz black branes. Phys. Rev. D **87**, 086005 (2013)
44. J.R. Sun, S.Y. Wu, H.Q. Zhang, Mimic the optical conductivity in disordered solids via gauge/gravity duality. Phys. Lett. B **729**, 177 (2014)
45. L.Q. Fang, X.H. Ge, X.M. Kuang, Holographic fermions in charged Lifshitz theory. Phys. Rev. D **86**, 105037 (2012)
46. L.Q. Fang, X.H. Ge, J.P. Wu, H.Q. Leng, Anisotropic Fermi surface from holography. Phys. Rev. D **91**, 126009 (2015)
47. X.H. Ge, S.J. Sin, S.F. Wu, Lower bound of electrical conductivity from holography. arXiv:1512.01917 [hep-th]
48. D. Roychowdhury, On anisotropic black branes with Lifshitz scaling. Phys. Lett. B **759**, 410 (2016). arXiv:1509.05229 [hep-th]
49. H. Liu, K. Rajagopal, U.A. Wiedemann, Wilson loops in heavy ion collisions and their calculation in AdS/CFT. JHEP **0703**, 066 (2007)
50. R. Baier, Y.L. Dokshitzer, A.H. Mueller, S. Peigne, D. Schiff, Radiative energy loss and p(T)-broadening of high energy partons in nunuclei. Nucl. Phys. B **484**, 265 (1997)
51. X.H. Ge, Y. Matsuo, F.-W. Shu, S.-J. Sin, T. Tsukioka, Density dependence of transport coefficients from holographic hydrodynamics. Progr. Theor. Phys. **120**, 833 (2008)
52. X.H. Ge, Y. Ling, Y. Tian, X.N. Wu, Holographic RG flows and transport coefficients in Einstein-Gauss-Bonnet-Maxwell theory. JHEP **1201**, 117 (2012)
53. S. Jain, R. Samanta, S.P. Trivedi, The shear viscosity in anisotropic phases. JHEP **1510**, 028 (2015)